\documentclass[usenatbib]{mn2e}

\usepackage{graphicx}
\usepackage{lscape}

\title{Discovery of X-ray Pulsations from the HMXB Source AXJ1749.1-2733}
\author[D. I. Karasev, S. S. Tsygankov, A. A. Lutovinov]{D. I. Karasev$^{1}$\thanks{E-mail:dkarasev@hea.iki.rssi.ru}, S. S. Tsygankov$^{1}$, A. A. Lutovinov$^{1}$\\
$^{1}$Space Research Institute, Profsoyuznaya Str. 84/32, Moscow 117997, Russia\\
}

\date{Accepted .... Received ....}
\begin{document}
\pagerange{\pageref{firstpage}--\pageref{lastpage}} \pubyear{2007}

\maketitle

\label{firstpage}

\begin{abstract}

We are reporting a discovery of X-ray pulsations from the source AX
J1749.1-2733 with the period of $\sim$132 s based on the XMM-Newton
data obtained in March 2007. The observed pulse profile has a
double-peaked structure with the pulse fraction of about $25-30$\%
in the 3-10 keV energy band. We have also found that a periodicity
with practically the same period has been detected from the source
by the IBIS telescope onboard the INTEGRAL observatory during an
outburst on Sept. 9, 2003 in the 20-50 keV energy band. Due to the
double-peaked pulse profile, there is an additional peak on both
periodograms of nearly $\sim66$ s, therefore we have also
investigated the possibility that the last value is the true pulse
period. The source spectrum obtained by the XMM-Newton observatory
in the soft energy band is being heavily absorbed
($N_H\simeq2\times10^{23}$ cm$^{-2}$) due to a strong intrinsic
absorption in the binary system that leads to the conclusion that AX
J1749.1-2733 is a new transient X-ray pulsar in the high mass X-ray
binary system.

\end{abstract}


\section{Introduction}

The source AX J1749.1-2733 was discovered on Sept. 19, 1996 by the
ASCA space observatory \citep{Sak02}. The object had not been
detected during previous observations of this region in 1995 even
with a higher exposure and was significantly detected again only in
September 1997 and March 1998, which reflects its transient nature.
Due to the faintness of the source emission ($1.5-6\times10^{-12}$
erg cm$^{-2}$ s$^{-1}$ in the $0.7-10$ keV energy band), it was
practically impossible to obtain good restrictions to its spectral
parameters, but some evidence for the strong absorption in the
system was mentioned \citep{Sak02}.

On Sept. 9, 2003 a short (with a duration of $\sim1$ day) outburst
from the source was observed by the INTEGRAL observatory in hard
X-rays with the peak flux of $\sim40$ mCrab in the 20-60 keV energy
band \citep{Greb04}. Detailed investigations of the source behavior
and its characteristics during the outburst allowed \citet{Greb07}
to classify AX J1749.1-2733 as a fast X-ray transient. During the
outburst, the source position in hard X-rays was resolved as RA=17h
49m 07s, Dec=-27\fdg 32\arcmin 38\arcsec\ (J2000) with the
uncertainty of 1.8\arcmin\ \citep{Sgu06}, that is coincident with
the source position revealed by the ASCA observatory -- RA=17h 49m
10s, Dec=-27\fdg 33\arcmin 14\arcsec (J2000) the uncertainty of
about 55\arcsec, \citet{Sak02}. Subsequent analysis showed that the
source had been significantly detected by INTEGRAL from time to time
in 2004-2007, but the typical value of its flux was about $5-8$
mCrab. Nevertheless, based on these data, \citet{Zur07} suggested
the presence in the system of the $\sim185$-day orbital period.
Observations of AX J1749.1-2733 performed in soft X-rays by the
SWIFT observatory in 2007, allowed \citet{Kong07} to improve its
localization to 3.8\arcmin\arcmin and revealed a strong absorption
($N_H\simeq1.9\times10^{23}$ cm$^{-2}$) in the source spectrum.

In this work, we reanalyzed the INTEGRAL data obtained during the
outburst in September 2003 and analyzed the XMM-Newton data obtained
in March 2007. In both data sets, strong periodic modulations of the
X-ray flux with the period of $\sim$132 s were found. The
preliminary results were reported by \citet{Karas07}.

\section{Observations and Data Analysis}

The source AX J1749.1-2733 was observed with the XMM-Newton
observatory on March 31, 2007. Our analysis was based in general on
the data from the EPIC-PN camera, but the data from the MOS
telescopes were used also for the test of the consistency of the
used spectral models. All data were processed with the Science
Analysis System (SAS)\footnote{http://xmm.esac.esa.int/sas/}. We
provided a standard method for the filtration of the proton flares
produced as a result of the interaction between soft protons in the
Earth's magnetosphere with the telescope. The source spectrum and
lightcurve were extracted from the circle with the radius of
14\arcsec\ around the source; the background spectrum and lightcurve
were extracted from a nearby circle region with the same radius. A
total exposure of XMM-Newton/PN observations was approximately 6 ks.

%***********************************************************************
\begin{figure}
\centerline{\includegraphics[width=8cm,bb=15 415 570 700,clip]{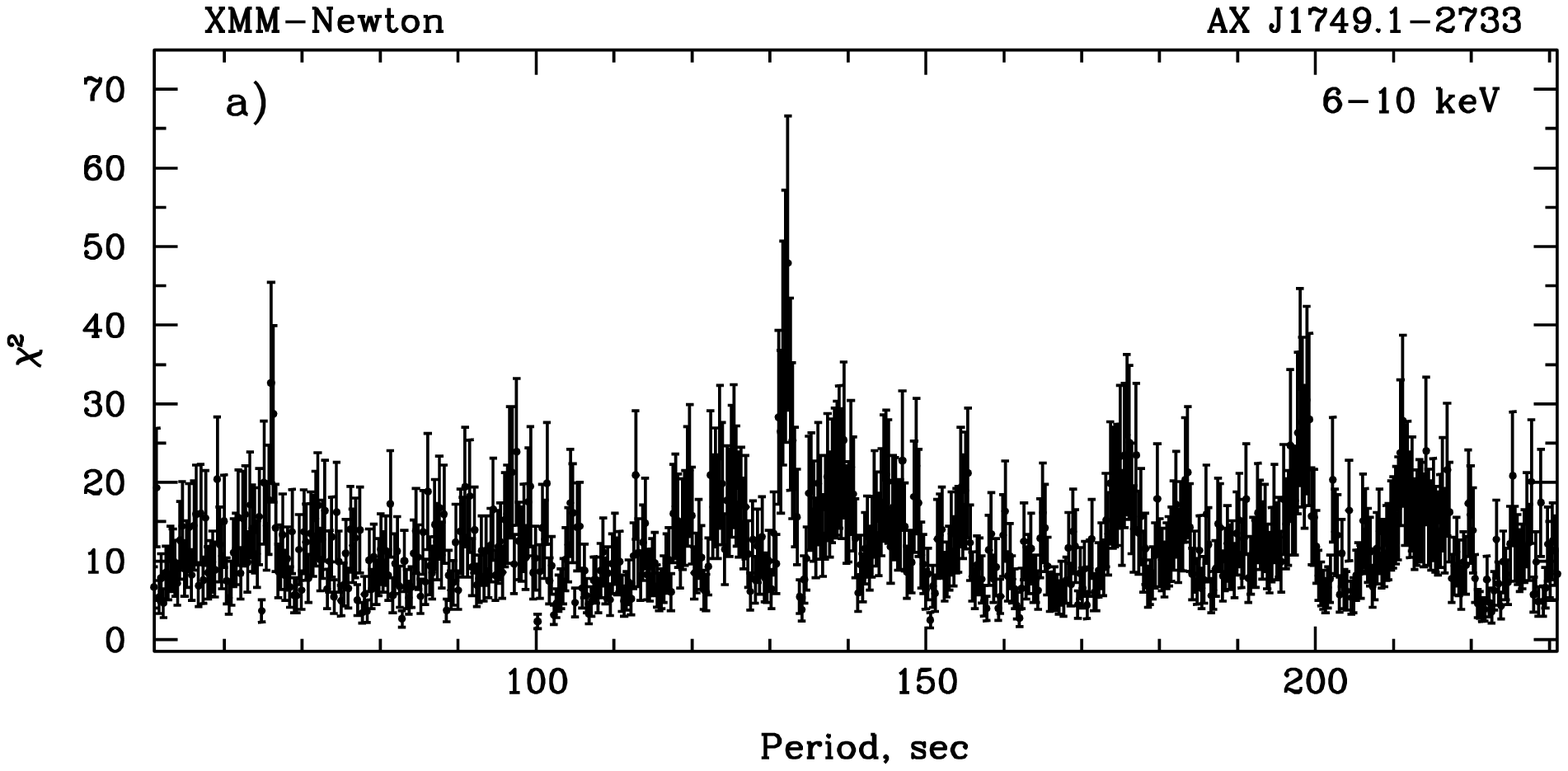}}
\centerline{\includegraphics[width=8cm,bb=15 415 570 700,clip]{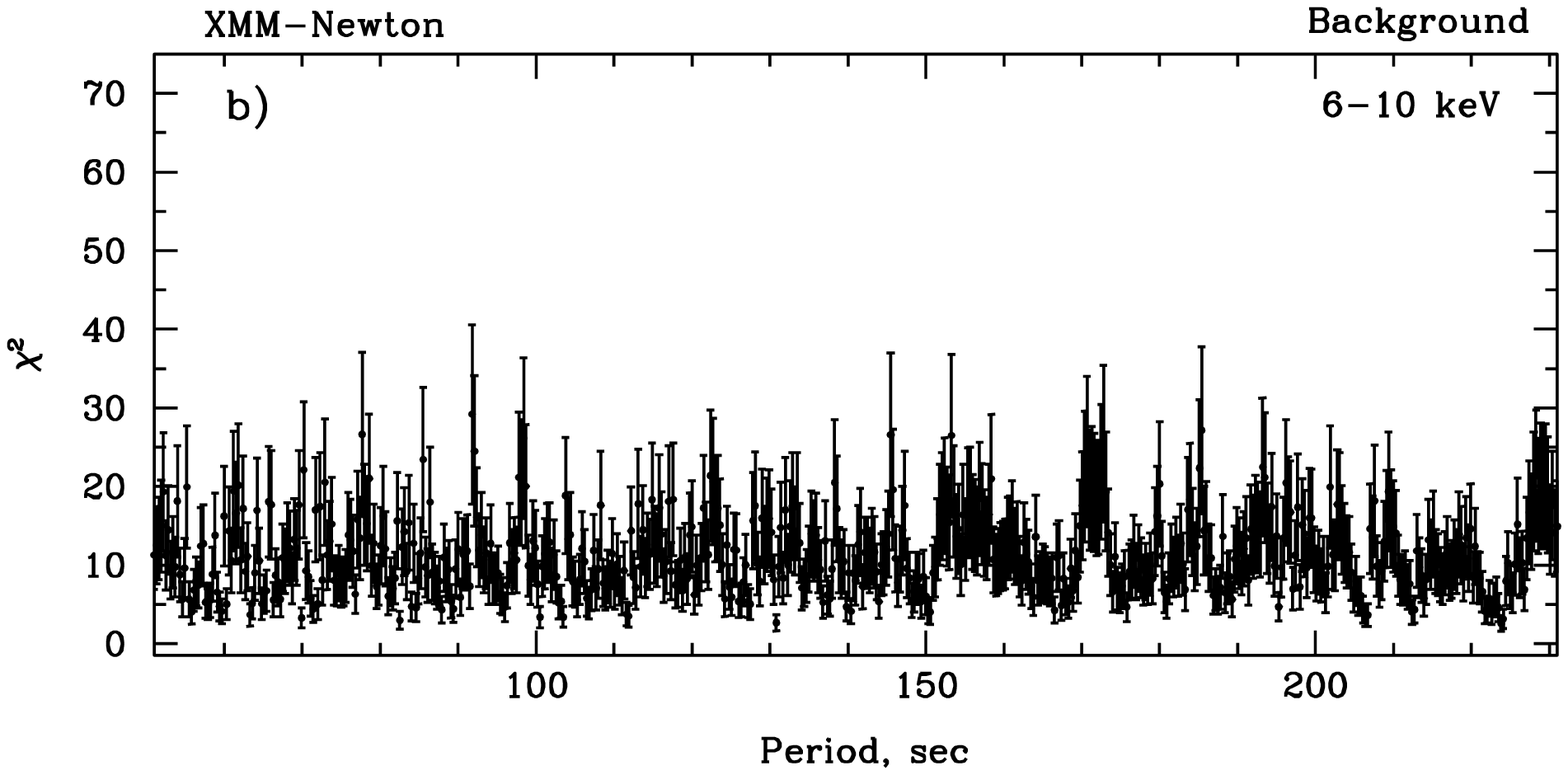}}

\caption{Periodograms of a background subtracted emission of AX
J1749.1-2733 (a) and background region (b) obtained with the
XMM-Newton observatory (the PN instrument) in the 6-10 keV energy
band.}\label{xmm.per}

\end{figure}
%***********************************************************************

As mentioned above, the source AX J1749.1-2733 had been
significantly detected by the IBIS telescope several times during
the four years of observation, but only once was its intensity
enough to detect the pulsations. The total exposure of the
INTEGRAL/IBIS observations during this outburst on Sept 9, 2003 was
about 50 ks. The INTEGRAL/IBIS data for the timing analysis were
processed using the standard software OSA 6.0, distributed by the
INTEGRAL Science Data Center, Versoix,
Switzerland\footnote{http://isdc.unige.ch}, and the software
designed and supported by the National Institute of Astrophysics in
Palermo,
Italy\footnote{http://www.pa.iasf.cnr.it/$\sim$ferrigno/INTEGRALsoftware.html}
\citep{Seg07}. For the spectral analysis, we used the software
developed at the Space Research Institute RAS, Moscow, Russia
(description of its main features can be found in \citet{Rev04} and
\citet{kris07}). The final timing and spectral analysis were
provided with the FTOOLS
package\footnote{http://heasarc.gsfc.nasa.gov/lheasoft}.

%*************************************************************************
\begin{figure}

\vspace{2mm}

\centerline{\includegraphics[width=9cm,bb=18 490 590 685,clip]{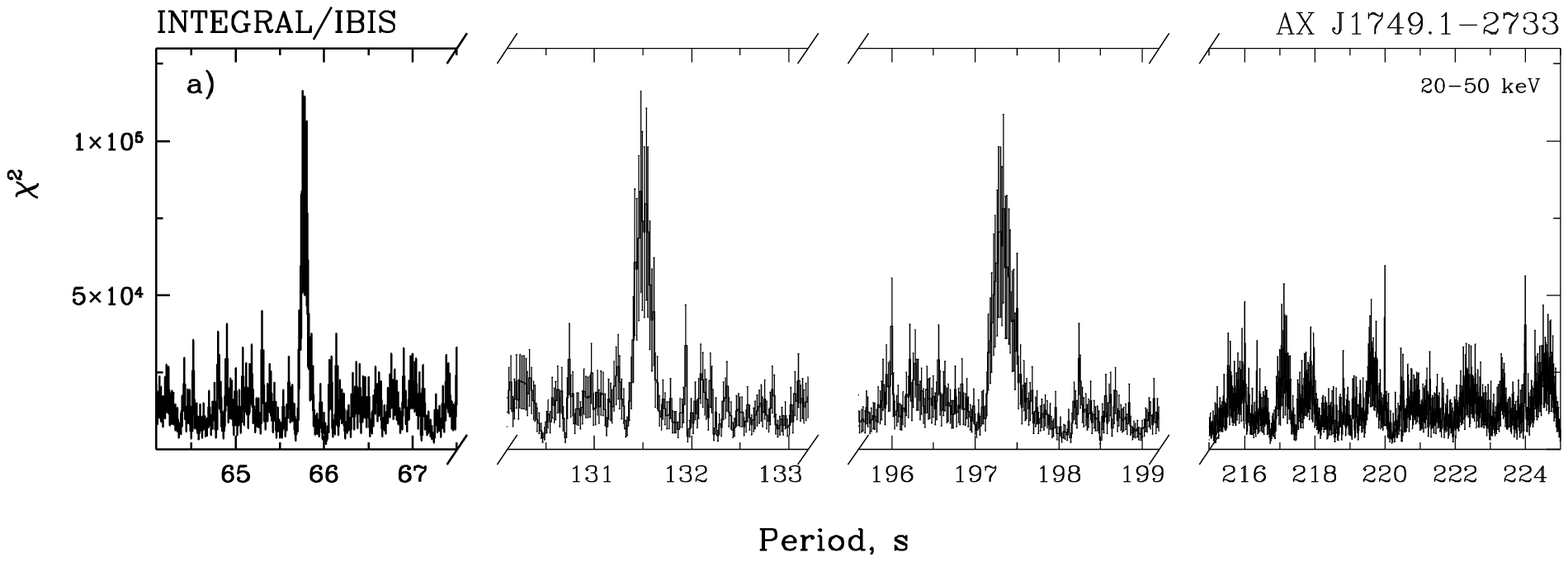}}

\centerline{\includegraphics[width=9cm,bb=18 490 590 685,clip]{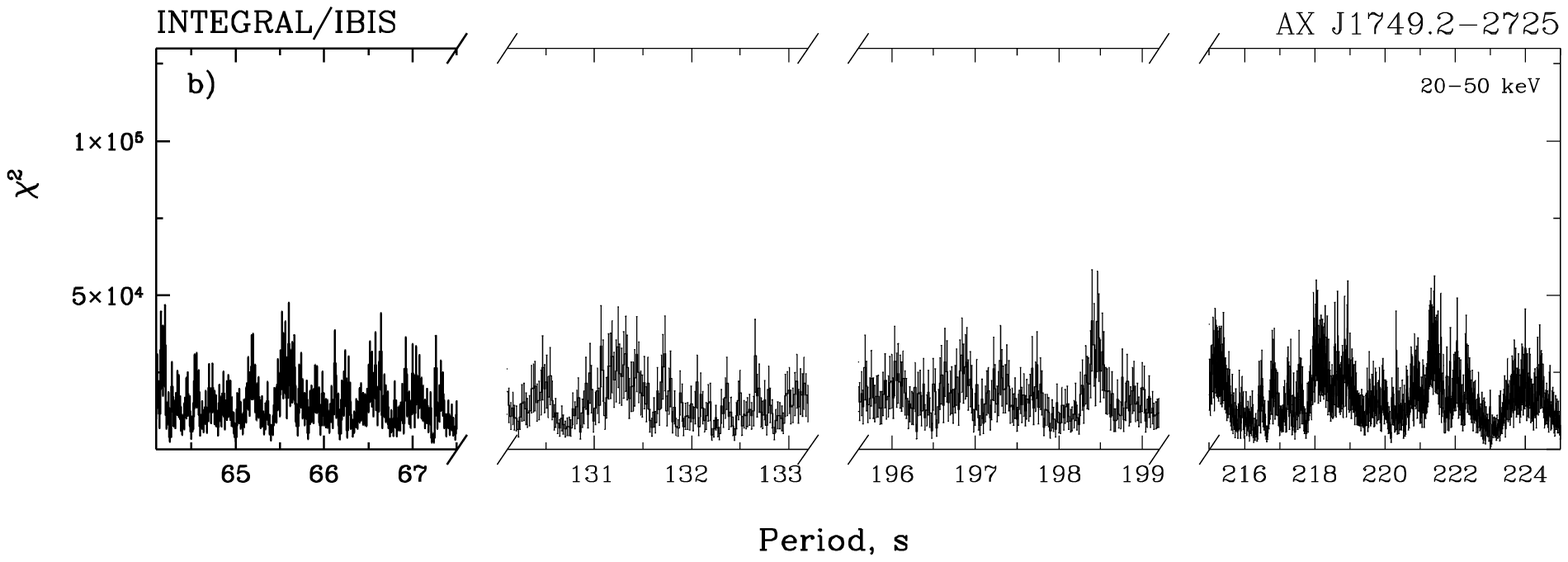}}

\caption{Periodograms of background subtracted emissions of AX
J1749.1-2733 (a) and AX J1749.2-2725 (b) obtained with the
INTEGRAL/IBIS data in the 20-50 keV energy band.}\label{int.per}

\end{figure}
%*************************************************************************

\section{Timing analysis}

Fig.\ref{xmm.per}a shows a periodogram of a background subtracted
emission of AX J1749.1-2733 obtained from the XMM-Newton/PN data in
the 6-10 keV energy band. The significant excess ($\sim6\sigma$) of
the $\chi^2$-distribution at $\sim$132 s can be clearly seen. Apart
from them, there are several other peaks of a lower intensity, which
are mostly prominent at the periods of $\sim$66 s (the significance
is $\sim3\sigma$) and $\sim$198 s, which corresponds to a half and
one and a half of the $\sim$132-s period, respectively. To clarify
that these excesses are not artificial and connected with the source
AX J1749.1-2733, we have also built a periodogram for emissions from
the nearby background region. No signals with the periods of 66, 132
or 198 s were detected on this periodogram (Fig.\ref{xmm.per}b).
This fact indicates that the origin of the detected periodicity is
connected with the source AX J1749.1-2733. The pulsations in the
source emission was checked also in hard X-rays ($>$20 keV) using
the data of the IBIS telescope onboard the INTEGRAL observatory. In
Fig.\ref{int.per}a, the periodogram of the 20-50 keV source emission
is shown near the periods of $\sim$66, $\sim$132, and $\sim$198 s.
Similarly to the XMM-Newton results, the signal can be clearly seen
again, but the significances of the $\sim66$ and $\sim132$ s peaks
in this case are roughly the same ($\sim15-16\sigma$).  The list of
possible pulse periods near 66 and 132 s measured with different
instruments is presented in Table 1. Errors (corresponding to
$1\sigma$) were determined by the bootstrap method from the analysis
of a large number of simulated light curves (see e.g. \citet{Tsy05}
for details).

%=================================================================
\begin{table}
%\begin{center}
{\bf Table 1. }{ List of AX J1749.1-2733 pulse periods}\\
\begin{tabular}{lll}
\hline\hline
Date & Observatory & Pulse Periods  \\
\hline
Sept 9, 2003 & INTEGRAL & $131.54\pm0.02$, $65.77\pm0.01$\\
March 31, 2007 & XMM-Newton & $131.95\pm0.24$, $66.05\pm0.15$\\
\hline
\end{tabular}
%\end{center}
\end{table}
%=================================================================

It is important to note that in a near vicinity to the studied
source there is located another transient X-ray pulsar AX
J1749.2-2725 with the pulse period of $\sim$ 220 s \citep{Tor98}.
Due to a small angular distance ($\sim$8\arcmin) between these
sources, there is a possibility of a ``photon-drifting'' from one
source to another in the processing of the data reduction of the
IBIS telescope (as its angular resolution is only 12\arcmin). To
check a possible influence of such an effect, we have done the same
timing analysis for AX J1749.2-2725. Fig.\ref{int.per}b shows the
resulting periodogram for this source. It can be seen that
significant peaks are detected neither at $\sim$ 132 s nor at $\sim$
220 s (note that no peak near 220 s can also be seen on the
AXJ1749.1-2733 periodogram, Fig. \ref{int.per}a). Most likely, the
pulsar AX J1749.2-2725 was in the ``switched-off'' state during the
INTEGRAL observations \citep{Greb07} and as such, did not affect our
results.

%*************************************************************************
\begin{figure}
\centerline{\includegraphics[width=8cm,bb=50 410 280 825,clip]{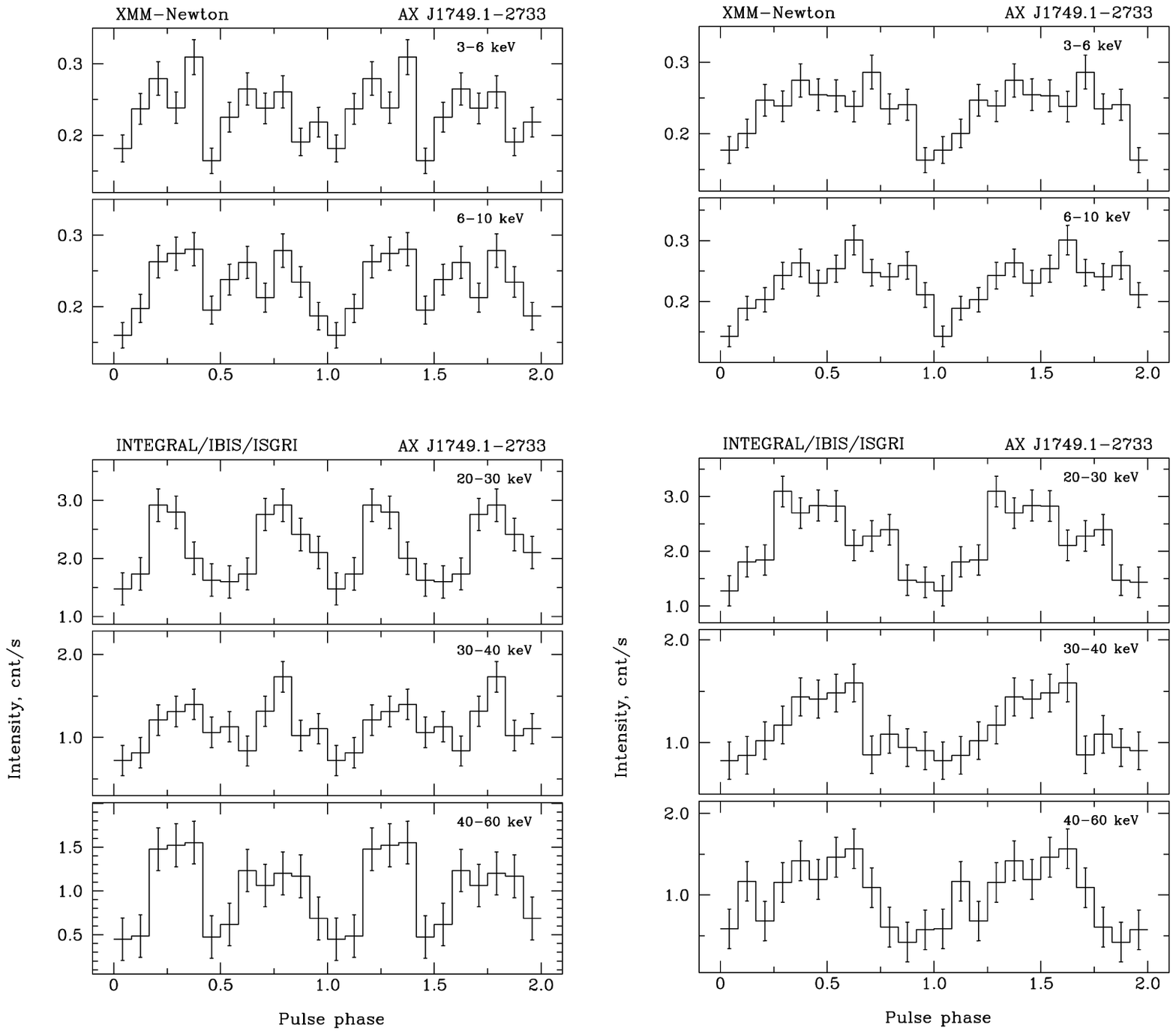}}

\caption{The AX J1749.1-2733 pulse profiles in the 3-6, 6-10, 20-30,
30-40 and 40-60 keV energy bands obtained with the XMM-Newton/PN
(two upper panels) and INTEGRAL/IBIS (three bottom panels)
instruments for the pulse periods of 131.95 s and 131.54 s,
respectively. Errors correspond to 1$\sigma$.}\label{pprof.2p}

\end{figure}
%*************************************************************************

Thus, the pulsations in the source AX J1749.1-2733 can be clearly
established. However, the question about the true value of its
period is still relevant. In Fig.\ref{pprof.2p}, the source pulse
phase curves obtained by XMM-Newton and INTEGRAL observatories and
folded with the period of $\sim$132 s are presented for several
energy bands: 3-6, 6-10, 20-30, 30-40 and 40-60 keV; due to slightly
different values of the pulse period, the data of both observatories
were convolved with the corresponding periods (see Table 1). The
source pulse profile builded in such a way has a clear double-peaked
structure at all energies that account for several peaks in the
periodograms with intervals divisible by $\sim66$ s.

%%%%%%%%%%%%%%%%%%%%%%%%%%%%%%
\begin{figure}
\centerline{\includegraphics[width=8.5cm,bb=85 520 500 710,clip]{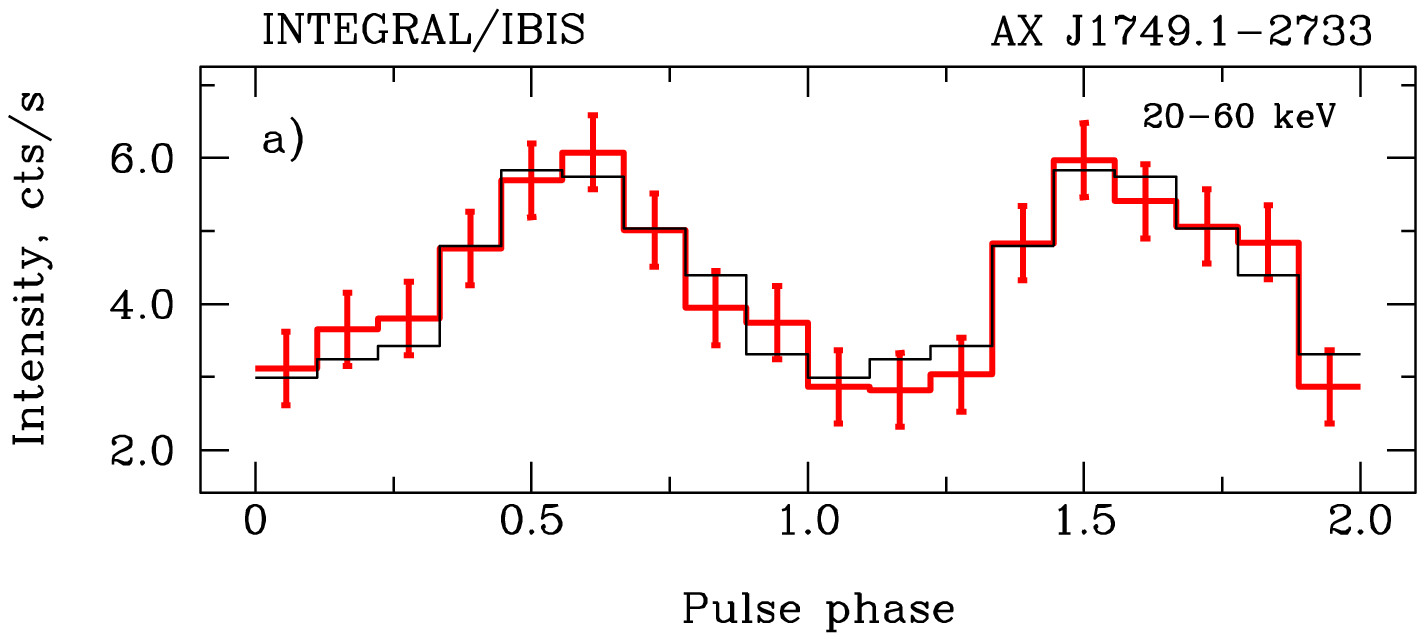}}

\centerline{\includegraphics[width=8.5cm,bb=85 520 500 710,clip]{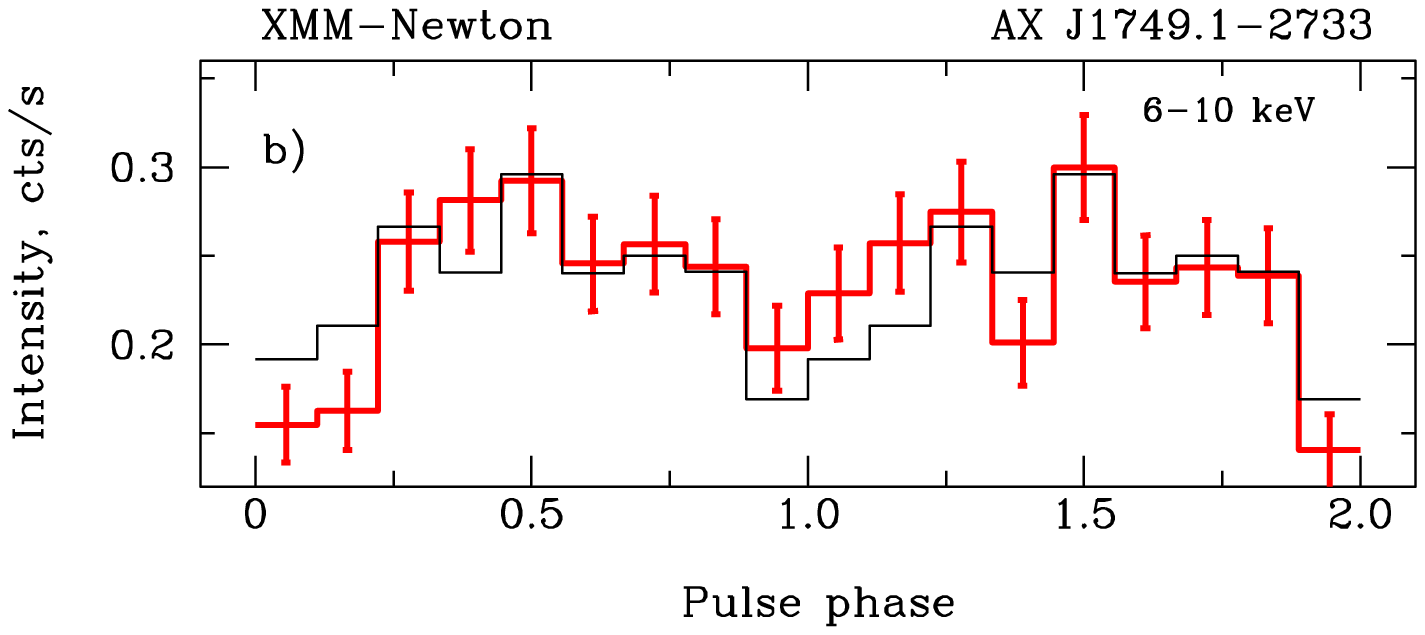}}

\caption{(a) Pulse profiles of AX J1749.1-2733 obtained with the
INTEGRAL/ISGRI data and folded with two different periods: 131.54 s
(thick line) and 65.77 s (thin line). The second one has been
plotted twice to demonstrate a discrepancy between profiles on the
time scale of 131.54 s. (b) The same, but the periods 131.95 s and
66.05 s have been obtained with the XMM-Newton/PN
data.}\label{pp.dif}

\end{figure}
%%%%%%%%%%%%%%%%%%%%%%%%%%%%%%%%%%%%%

To clarify which of pulse periods is true, we used the following
procedure: the source pulse profile folded with the period of 131.54
s and the doubled source pulse profile folded with the period of
65.77 s obtained from the INTEGRAL/IBIS data in the 20-60 keV energy
band were plotted in the same figure and compared between themselves
(Fig.\ref{pp.dif}a). The test clearly demonstrates that there are no
significant differences between the pulse profiles obtained for
these two periods, which might be due to a relatively faint emission
from the source and large uncertainties in the observed count rates.
To formalize this test, we used a simple criteria in the form of
$\chi^2=\sum_{i=1}^{k} \frac{(X_i-\mu_i)^2}{\sigma_i^2}$, where
$X_i$ is the count rate in the $i$th bin of the 132 s folded light
curve, $\mu_i$ is the count rate in the $i$th bin of the 66 s folded
light curve, which is being used here as a model; and $\sigma_i$ is
the correspondence uncertainty. A summation is being done only on
the half of bins (which are statistically independent) of the 132 s
folded light curve (in our case, $k=9$). The application of this
test to the INTEGRAL/IBIS data leads to the value of 6.8 for 9
d.o.f., which means that the difference is insignificant. This
result reflects the practically equal significance of the 66-s and
132-s peaks at the correspondence periodogram (Fig.\ref{int.per}a).
The same procedure was used for the XMM-Newton data obtained in the
6-10 keV energy band. We folded them again with the corresponding
periods (see Table 1) and found that the difference between the
pulse profiles in this energy band is more significant
(Fig.\ref{pp.dif}b). Its value (22 for 9 d.o.f.) corresponds to the
probability of $\sim0.01$ that this difference is causal. Note that
the suggested procedure is formally equivalent to the comparison of
the odd and even 66 s segments on the 132 s folded light curve.
Thus, the use of the XMM-Newton data brings us to the conclusion
that $\sim132$ s is the probable pulse period of the source AX
J1749.1-2733.

During the XMM-Newton observations, the pulse period was slightly
longer than during the INTEGRAL ones. Due to a relatively high
uncertainty of the XMM-Newton period measurements, this difference
is not very significant ($\sim2\sigma$), but if we suggest that it
be connected with the real deceleration of the neutron star
rotation, the formal ${\dot P/P}$ will be about $10^{-3}$ yr$^{-1}$,
which could be consistent with the typical rates of the pulse period
change for X-ray pulsars (see e.g. \citet{Bild97,Lut94}).

\begin{figure}
\centerline{\includegraphics[width=8cm,bb=30 280 515 690,clip]{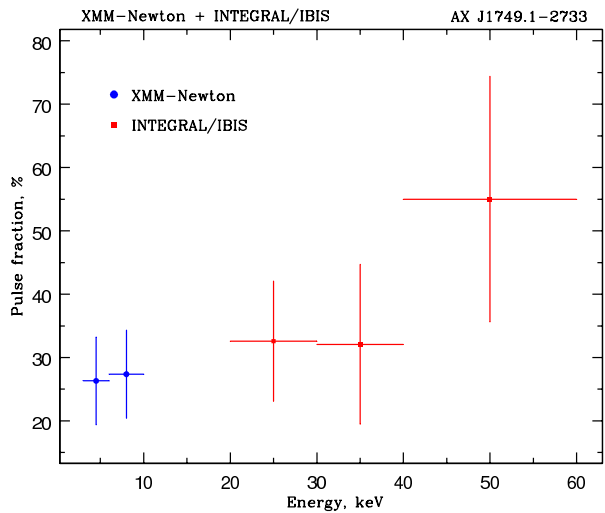}}

\caption{Pulse fraction dependence on the energy for the pulse
period of 131.95 and 131.54 s for the XMM-Newton (circles) and
INTEGRAL (squares) observatories, respectively. }\label{ppfrac}

\end{figure}

For all the energy bands (Fig.\ref{pprof.2p}), we calculated the
pulse fraction, which is defined as
$PF=(I_{max}-I_{min})/(I_{max}+I_{min})$, where $I_{min}$ and
$I_{max}$ are background-corrected count rates at the pulse profile
minimum and maximum. Fig.\ref{ppfrac} shows the energy dependence of
this value for both observatories. It is clear that the pulse
fraction is relatively constant in the energy band from $\sim3$ to
$\sim40$ keV at the level of $\sim25-30$\% and slightly increases up
to $\sim50$\% in hard X-rays ($>40$ keV), which is typical for X-ray
pulsars (see e.g. \citet{Tsyg07}).

%***************************************************************************
%*****************************************************************************

\section{Spectral Analysis}

It is important to note that the INTEGRAL and XMM-Newton data were
obtained not only in different epochs but also in different source
states --- INTEGRAL had observed the source during the outburst in
September 2003 and XMM-Newton had observed the source in March 2007
most likely in a quiescent state (see below). Thus, the combined
spectral analysis is possible only assuming the unchanged shape of
the source spectrum. However, as it was shown by \citet{Greb07}, the
source spectrum was changing significantly even during the outburst,
all the more, it can be different in the quiescent state. Therefore,
here we are providing the results of the spectral analysis
separately for the XMM-Newton data and only for the rough estimation
of the general spectral shape and its parameters in a wide energy
band where the XMM-Newton and INTEGRAL data were combined using
relative normalization.

%**********************************************************
\begin{figure}
\centerline{\includegraphics[width=9cm,bb=45 195 585 700]{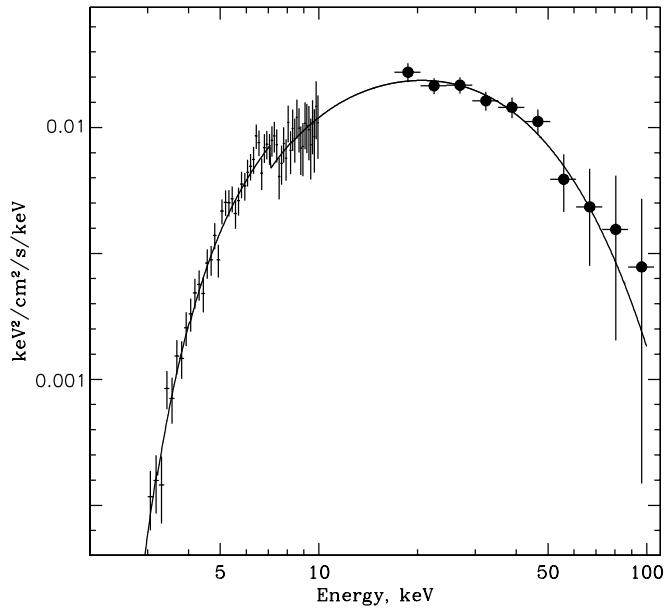}}

\caption{Broadband energy spectrum of AXJ1749.1-2733 in the 3-100
keV energy band; XMM-Newton (crosses) and INTEGRAL (circles) data
were combined using the relative normalization.}\label{spec}

\end{figure}
%********************************************************

The source spectrum obtained with XMM-Newton/PN can be well
described by a simple powerlaw with the photon index
$\Gamma=1.17^{+0.23}_{-0.06}$, the photoelectric absorption
$N_H=21.1^{+2.7}_{-1.3} \times10^{22}$ cm$^{-2}$ and the unabsorbed
source flux of $\sim1.8\times10^{-11}$ erg cm$^{-2}$ s$^{-1}$ in the
$3-10$ keV energy band (note that the use of the XMM-Newton/MOS data
gives approximately the same results).

The source spectrum in the soft X-ray band ($2-10$ keV) had
previously been measured with the XRT telescope of the SWIFT
observatory \citep{Rom07, Kong07} and the ASCA observatory
\citep{Sak02}. It was also approximated by the powerlaw model with
the photoelectric absorption. The values of $N_H$ comply with the
results of our analysis, but the values of $\Gamma$ are different,
which can be the result of the spectral shape variations similarly
as it was found by \citet{Greb07} for the outburst. On the other
hand, it is necessary to note that uncertainties of the photon index
measurements with the SWIFT and ASCA observatories are very large
and, nominally speaking, our results fall within the corresponding
error boxes.

A rough estimation of the interstellar absorption in the direction
to the source using the $N_H$ map \citep{Di90} gives the value of
$N_H\simeq1.63\times10^{22}$ cm$^{-2}$, which is more than one order
of magnitude lower than the value obtained from our spectral
analysis. Based on these measurements and previous results of the
ASCA and SWIFT observatories, we can undoubtedly claim that AX
J1749.1-2733 is a heavily absorbed source, which is probably
connected with the massive counterpart in the binary system (likely
a supergiant or a giant).

The formal approximation of the XMM-Newton and INTEGRAL data
simultaneously with the "standard" pulsars model \citep{Whi83}
modified by the photoelectric absorption gives the photon index of
$\Gamma=1.03^{+0.18}_{-0.24}$, the cutoff energy $E_{cut}=
7.1^{+0.6}_{-0.9}$ keV, the folding energy $E_{cut}=
19.8^{+1.9}_{-2.6}$ keV, the absorption column
$N_H=(20.2^{+1.0}_{-1.9}) \times10^{22}$ cm$^{-2}$ with the
corresponding value of the reduced $\chi^2$=0.94 (Fig.\ref{spec}).
Note, that the obtained values and the spectrum shape are typical
for X-ray pulsars (see e.g. \citet{Fil05}) despite the roughness and
a certain randomness of such a combination of the XMM-Newton and
INTEGRAL data. The broadband $3-100$ keV X-ray fluxes corresponding
to the XMM-Newton and INTEGRAL epochs of observations differ
approximately by six times: $\sim4.9\times10^{-11}$ ergs
cm$^{-2}$s$^{-1}$ versus $\sim2.8\times10^{-10}$ ergs
cm$^{-2}$s$^{-1}$, respectively. The last value corresponds to the
results of \citet{Greb07} obtained from a combined analysis of the
INTEGRAL/JEM-X and INTEGRAL/IBIS data.

\section{SUMMARY}

We are reporting a discovery of X-ray pulsations from AX
J1749.1-2733 by the INTEGRAL and XMM-Newton observatories with the
periods of 131.54 s in September 2003 and 131.95 s in March 2007,
respectively. The pulse profile has a double-peaked shape in a wide
energy band ($3-60$ keV) and the pulse fraction slightly increases
with the energy from $\sim25-30$\% to $\sim50$\%.  The source
spectrum can be described by the powerlaw model with the
photoelectric absorption, the value of which is much higher than the
interstellar one in the direction to the source. Thus, we can
conclude that AX J1749.1-2733 must be a new transient X-ray pulsar
in the high-mass X-ray binary system with a strong intrinsic
absorption like IGR J16465-4507 \citep{Lut05} or AX J1841.0-0536/IGR
J18410-0535) \citep{Hal04}.

It is interesting to compare the AX J1749.1-2733 position with the
position of different classes of binary sources on the Corbet
$P_{orb}-P_s$ diagram \citep{Corb86}. According to the relation
between the orbital period of $\sim185$ days suggest by
\citet{Zur07} and the source pulse period of 132 s, this source can
belong to the class of Be-systems with the orbital eccentricity of
$\sim0.3$ (see eq.[1] of \citet{Corb86}). But the measured value of
the intrinsic absorption in the binary system is repeatedly high and
larger than it is usually observed in Be-systems and is typical for
supergiant systems.

Note that the INTEGRAL and XMM-Newton observations are separated in
time by an approximately integer of the suggested orbital periods
($\sim7$). Therefore we checked the possibility that the XMM-Newton
observations might have been performed also during an outburst.
Using the moment of the 2003 outburst maximum MJD52891.65
\citep{Greb07} and the orbital period value of 185 days, we
determined the moment of the 2007 outburst maximum as MJD 54186.65.
The XMM-Newton observations were performed $\sim3.8$ later. As the
2003 outburst duration was about 1 day \citep{Greb07}, it is very
unlikely, for the suggested orbital parameters, that the XMM-Newton
observations were performed also during an outburst. Taking into
account the measured ratio of the INTEGRAL and XMM-Newton fluxes (6
times), it is possible to combine the obtained results to suppose
that the orbital period is slightly longer, $\sim185.5$ days.
However, this supposition requires an additional study.

This question as well as the one about the nature of the system will
be addressed in detail in a separate paper with data of the AX
J1749.2-2733 special observations performed with the Russian-Turkish
Telescope (RTT-150) in August 2007 (Karasev at al., 2008, in
preparation).

\section {ANKNOWLEGEMENTS}

We would like to thank E.M.\,Churazov for developing the IBIS data
analysis algorithms, providing the software and the discussion of
the timing analysis results. We extend our gratitude to
S.A.\,Grebenev and R.A.\,Sunyaev for the discussion of the results
and for the useful remarks. We would also like to pay tribute to the
very useful comments of an anonymous referee, which allowed us to
significantly improve the paper. We used data from the archive of
the Goddard Space Flight Center (NASA), the Integral Science Data
Centre (Versois, Switzerland), and the Russian Science Data Center
for INTEGRAL (Moscow, Russia). This work was supported by the
Russian Foundation for Basic Research (Project No. 07-02-01051), the
Presidium of the Russian Academy of Sciences (The Origin and
Evolution of Stars and Galaxies Program), and the Program of the
President of the Russian Federation for the Support of Scientifc
Schools (Project No. NSh-1100.2006.2). A.A. Lutovinov would like to
thank the Russian Science Support Foundation as well. AL and ST thank 
International Space Science Institute (ISSI, Bern, Swiss) for the 
hospitality and partial support.

\label{lastpage}

\end{document}